\documentclass[letterpaper,twocolumn,10pt]{article}
\usepackage{usenix,epsfig,amsmath,url,subfig}

\begin{document}

\date{}

\title{\Large \bf Green: Towards a Pollution-Free Peer-to-Peer Content Sharing Service}

\author{
{\rm Ruichuan Chen$^\dag$, Eng Keong Lua$^\ddag$, Zhuhua Cai$^\dag$, Jon Crowcroft$^\S$, Zhong Chen$^\dag$}\\
$^\dag$Peking University, China; $^\ddag$Carnegie Mellon University,
US; $^\S$University of Cambridge, UK
} 

\maketitle

\thispagestyle{empty}

\subsection*{Abstract}

Peer-to-Peer (P2P) content sharing systems are susceptible to the
content pollution attack, in which attackers aggressively inject
polluted contents into the systems to reduce the availability of
authentic contents, thus decreasing the confidence of participating
users.

In this paper, we design a pollution-free P2P content sharing
system, \emph{Green}, by exploiting the inherent content-based
information and the social-based reputation. In Green, a content
provider (i.e., creator or sharer) publishes the information of his
shared contents to a group of content maintainers self-organized in
a security overlay for providing the mechanisms of redundancy and
reliability, so that a content requestor can obtain and filter the
information of his requested content from the associated
maintainers. We employ a reputation model to help the requestor
better identify the polluted contents, and then utilize the social
(friend-related) information to enhance the effectiveness and
efficiency of our reputation model. Now, the requestor could easily
select an authentic content version for downloading. While
downloading, each requestor performs a realtime integrity
verification and takes prompt protection to handle the content
pollution. To further improve the system performance, we devise a
scalable probabilistic verification scheme.

Green is broadly applicable for both structured and unstructured
overlay applications, and moreover, it is able to defeat various
kinds of content pollution attacks without incurring significant
overhead on the participating users. The evaluation in massive-scale
networks validates the success of Green against the content
pollution.

\section{Introduction}

Peer-to-Peer (P2P) content sharing systems have experienced an
explosive growth since their inception in 1999, and now dominate
large fractions of both the Internet users and traffic
volume~\cite{KumarYBRR06}. However, due to the decentralized and
unauthenticated nature, each participating user has to manage the
potential risks involved in the application transactions without
adequate experience and knowledge about other users.

Recently, many measurement studies reported that content pollution
is pervasive in P2P content sharing systems, e.g., more than $50\%$
of the copies of popular songs are polluted in
KaZaA~\cite{LiangKXR05}. In a typical content pollution attack, the
polluters first collectively tamper with the target content,
degrading its quality or rendering it unusable. Then, they inject a
massive number of tampered (i.e., polluted) contents into the
system. Unable to distinguish between authentic contents and
polluted contents, genuine users download the polluted contents
unsuspectingly into their own shared folders, from which other users
may download later without knowing that these contents have been
polluted, thus passively contributing to the dissemination of
pollution. As a result, the polluted contents spread through the
whole system at extraordinary speed. Such content pollution has the
detrimental effects of reducing the availability of the original
authentic content, and ultimately, decreasing the confidence of
users in the P2P content sharing system.

In general, the simple digital signature scheme could be used to
defend against content pollution in many Internet applications.
Nevertheless, due to the lack of centralized trusted authorities in
P2P content sharing systems, the applicability of the signature
scheme is questionable. So far, many pollution defenses have been
further proposed, and most of them are built upon reputation
models~\cite{CostaA07,CostaSAA07,CurtisSS04,DamianiVPSV02,KamvarSG03,WalshS06,XiongL04}.
These models utilize the information derived from the participating
users' feedback; however, they are penalized by the lack of reliable
user cooperation, and cannot defend against one concrete pollution
attack: identifier corruption (described in
section~\ref{subsec:corruption}). In this paper, we design
\emph{Green}, a pollution-free P2P content sharing system. Our main
motivation is to provide a total defense against various kinds of
existing pollution attacks by exploiting not only the traditional
reputation-based information but also the inherent content-based
information and social-based information.

Firstly, in Green, a content provider (i.e., creator or sharer)
publishes his shared contents' information to a group of content
maintainers self-organized in a security overlay for providing the
mechanisms of redundancy and reliability. This allows a content
requestor to obtain the authentic information of his requested
content by first looking up in the overlay for the associated
maintainers, and then executing a proactive filtering process to
filter out the malicious information corrupted by malicious and
compromised maintainers.

Secondly, after the proactive information filtering, a requestor
resorts to the reputation-based information to better identify the
polluted contents. Here, the requestor collects the past vote
histories of the users who have voted the requested content, so that
he could utilize the statistical correlation between his local vote
history and each of these collected vote histories to compute the
reputation score of any version of the requested content.

Thirdly, in order to enhance the effectiveness and efficiency of our
basic reputation model, a requestor further utilizes the social
(friend-related) information. Specifically, the requestor uses his
friends' vote histories to extend the local vote history reliably.
By considering these extended vote histories, each requestor is able
to perform the reputation computation more accurately; furthermore,
if many friends have already voted the requested content, the
requestor could use these associated friends' votes to efficiently
estimate the reputation score.

Finally, according to the authentic content-based information as
well as the social-based reputation information, the requestor
selects an authentic version of his requested content for
downloading. In Green, we allow a requestor to verify the integrity
of the requested content while downloading and take prompt
protection actions to handle content pollution attacks. To reduce
the verification overhead, we devise a scalable probabilistic
verification scheme in which each requestor has the capability of
flexibly adjusting the false positive rate of integrity verification
according to the tradeoff between integrity assurance and
verification overhead.

Green is broadly applicable for both structured and unstructured
overlay applications. Moreover, it is able to defend against various
kinds of content pollution attacks without incurring significant
overhead on the participating users. We implemented a prototype
system, and evaluate its performance on two massive-scale testbeds
with realistic network traces. The evaluation results illustrate
that Green can effectively and efficiently defend against content
pollution in various scenarios.

The rest of this paper is organized as follows. We specify the
system model and terminology in section~\ref{sec:model}, followed by
a description of threat model in section~\ref{sec:attack}. The
details of our Green system are elaborated in
section~\ref{sec:framework}. We then analyze the defense capacity
and maintenance overhead of Green in section~\ref{sec:analyze}.
Section~\ref{sec:evaluation} presents the experimental design and
discusses the evaluation results. Finally, we give an overview of
related work in section~\ref{sec:related}, and conclude this paper
in section~\ref{sec:conclusion}.

\section{System Model and Terminology}
\label{sec:model}

Current P2P overlay networks can generally be grouped into two
categories~\cite{LuaCPSL05}: structured overlay networks, e.g.,
Chord~\cite{StoicaMKKB01} and Pastry~\cite{RowstronD01}, that
accurately build an underlying topology to support rapid searching,
and unstructured overlay networks, e.g.,
Gnutella-like~\cite{Gnutella} networks, that merely impose loose
structure on the topology. In Green, for providing the mechanisms of
redundancy and reliability, we choose a structured P2P overlay
network as the underlying structure. Without loss of generality, we
utilize Chord as the dedicated underlying P2P overlay network, and
we can also conveniently utilize another structured overlay as an
alternative. Nowadays, the unstructured overlay applications are
prevalent on the Internet; to illustrate the broad applicability, we
will further describe how to deploy Green into unstructured overlays
in section~\ref{subsec:unstructured}.

Generally, a P2P content sharing system is composed of contents and
users. To elaborate our design clearly, we introduce some related
terminology about contents and users in the following two sections,
respectively.

\subsection{Content-Related Terminology}

We refer to a specific content (e.g., a document, video or song)
existing in the P2P content sharing system as a \emph{title}. A
given title can generally have multiple different \emph{versions},
each of which is published by a group of users. For instance, a
large number of versions of the same video could be produced by
various rippers/encoders, each of which may create a slightly
different version. Specifically, each version has an
\emph{identifier}, which is typically a hash value of the associated
data (and metadata) of the version; therefore, modifications of data
(and metadata) no matter faithfully or maliciously could also
produce additional different versions. Finally, each version may
have many \emph{copies} in the system due to the fact that
participating users could lookup and download various versions of
different titles from each other.

\subsection{User-Related Terminology}

In general, there is no centralized infrastructure in P2P content
sharing systems, and each participating user can play one or several
of the following three roles: \emph{provider}, \emph{requestor} and
\emph{maintainer}.

A provider publishes the information of his shared copies into the
P2P content sharing system, and informs the associated maintainer(s)
of building up or refreshing the corresponding index structure. On
the other hand, a requestor who wants to obtain a copy of a specific
title can utilize the underlying overlay to route a query towards
the maintainer(s) associated with the requested title, then these
maintainer(s) should respond with a list of corresponding index
records. Once receiving the index records, the requestor could
filter out malicious records and select one version to download from
the associated providers in parallel.

\section{Threat Model}
\label{sec:attack}

To better understand the characteristics of pollution attack and
design a more effective pollution defense, we discuss two attack
forms of content pollution: \emph{decoy insertion}~\cite{LiangKXR05}
and \emph{identifier corruption}~\cite{BenevenutoCVAAM06}.

\subsection{Decoy Insertion}

Decoy insertion is a common pollution mechanism utilized by
polluters in today's P2P content sharing systems. Here, the
polluters target a specific title, and then manufacture
\emph{decoys} (i.e., corrupted versions with the same metadata but
different identifiers) for the title in various ways, e.g., for
media contents, inserting advertisements, cutting the duration, and
replacing parts of the actual data with undecodable white noise.
Afterwards, the polluters insert a massive number of decoys of the
target title into the system in order to reduce the availability of
authentic versions of the title severely.

In general, when a content requestor searches for a title, the
associated maintainer(s) should group the information of the
requested title's available copies into different versions, and then
return the grouped result back to the requestor. Unable to
distinguish between authentic versions and polluted versions, the
requestor would select the version with the largest number of copies
to download; unfortunately, under collective decoy insertion
attacks, this selected version may be polluted.

\subsection{Identifier Corruption}
\label{subsec:corruption}

In a P2P content sharing system, each version is associated with an
identifier, which is typically generated by applying a hash function
to the data (and metadata) of the version. The generated version
identifier can be used by content requestors to identify different
versions of a specific title, and moreover, this identifier is also
important for maintainers to group the information of the requested
title's available copies into different versions.

Due to the feature of hash function, it is usually assumed that
different data (and metadata) would generate different identifiers.
However, it is still possible to have two actually different
versions with the same identifer, especially for the consideration
of efficiency, when some widely used P2P content sharing systems
adopt weak hash functions based only on a fraction of the data (and
metadata) of a version, e.g., the UUHash~\cite{UUHash} function
employed by KaZaA~\cite{KaZaA}. So now, the polluters could analyze
the adopted hash function, and corrupt the fraction of data (and
metadata) that are not used without altering the identifier of the
corrupted version.

Under identifier corruption attacks, when a requestor searches for a
title, he can receive the grouped information of the requested
title's versions from the maintainers, where each version is
associated with an identifier. As usual, the requestor selects one
version and downloads different blocks of the selected version from
different associated providers in parallel. However, if a downloaded
block is a corrupted part from the changed version, the entire
downloaded version would be polluted.

\section{Design of Green}
\label{sec:framework}

\begin{figure}[tbp]
    \centering
    \includegraphics[width=0.39\textwidth]{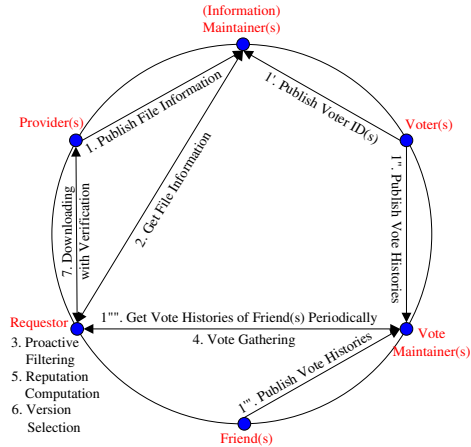}
    \caption{System illustration. Note that, each user practically plays one or several of these roles simultaneously.} \label{fig:stages}
\end{figure}

In different P2P content sharing systems, each title may be
associated with a \emph{topic}, a \emph{file} or a \emph{keyword},
depending on different system designs. Without loss of generality,
we use ``file'' to replace ``title'' hereafter for clarity.

In Green, a provider publishes the information of his shared files
to a group of maintainers self-organized in a structured overlay
(section~\ref{subsec:publication}), so that a requestor could
obtain/filter the information of the requested file from the
associated maintainers (section~\ref{subsec:filtering}). Afterwards,
the requestor utilizes a social-based reputation model to identify
the polluted versions of his requested file
(section~\ref{subsec:reputation}), and then selects an authentic
version for downloading (section~\ref{subsec:select}). While
downloading, we employ a realtime verification mechanism to help the
requestor take prompt actions to handle content pollution
(section~\ref{subsec:downloading}). Finally, we discuss how to
deploy our design in unstructured overlays
(section~\ref{subsec:unstructured}). In particular, the abstract
execution process of Green is illustrated in
Figure~\ref{fig:stages}, and the notation that we use to simplify
the description of our design can be found in
Table~\ref{tab:notation}.

\begin{table}[tbp]
    \centering
    \caption{Design Notation}
    \label{tab:notation}
    \begin{tabular}{|r|l|}
        \hline \textbf{Symbol} & \textbf{Meaning} \\
        \hline $F$             & File \\
               $V_i$           & The $i^{th}$ version of the requested file \\
               $VI_i$          & The identifier of the version $V_i$ \\
               $D$             & The digest of a data block \\
               $DL_i$          & The digest list of the version $V_i$ \\
               $U$             & User \\
               $UI$            & User identifier \\
               $P$             & Provider \\
               $PI$            & Provider identifier \\
               $PIL_i$         & The provider identifier list of the version $V_i$ \\
               $R$             & Requestor \\
               $M_i$           & The $i^{th}$ maintainer of a file \\
               $OI$            & Voter identifier \\
               $OIL_i$         & The voter identifier list of the version $V_i$ \\
               $O_{ij}$        & The $j^{th}$ voter who has voted the version $V_i$ \\
               $o_{ij}$        & The vote on $V_i$ given by the voter $O_{ij}$ \\
               $OH_X$          & The vote history of the user $X$ \\
               $OM_i$          & The $i^{th}$ vote maintainer for a user \\
               $F_i$           & The $i^{th}$ friend of the requestor \\
        \hline
    \end{tabular}
\end{table}

\subsection{Content Publication}
\label{subsec:publication}

As described in section~\ref{sec:model}, the underlying network
structure is assumed to be the Chord overlay, so we utilize SHA1 to
assign each user and file an identifier. The \emph{user identifier}
of a participant is chosen by hashing the participant's IP address,
while a \emph{file identifier} is produced by hashing the filename.
Recently, many users in P2P content sharing systems reside behind
network address translators (NATs), hence a number of participating
users may share the same public IP address. Moreover, we could not
guarantee that each user's IP address stays constant in a dynamic
P2P environment. That is, we are not able to distinguish between
different users solely by their IP addresses. To solve the NAT and
mobility problems, we may adopt another kind of unique and constant
identifiers (e.g., HIP-based identifiers~\cite{RFC5201}) to replace
the IP-based identifiers, which is not the focus of this paper.

Both the user identifiers and file identifiers are ordered in an
identifier circle. Specifically, the \emph{information} of a file
$F$ is assigned to the first user whose identifier is equal to or
follows the $F$'s file identifier in the identifier circle. This
user is called the successor of file $F$, denoted by $successor(F)$.
Such structure tends to balance the load on the system, since each
user maintains the information of roughly the same number of files;
even if a user maintains the information of a popular file, several
existing techniques~\cite{GodfreyLSKS04,RaoLSKS03} have provided the
solutions of load balancing.

In Green, a provider $P$, who wants to publish a specific file $F$,
first divides the file $F$ into $b$ data blocks, and then, he
computes the digest of each block to construct a \emph{digest list},
i.e., $\{D_i\}_{i=1}^{b}$. Here we could further utilize the
homomorphic hashing technique~\cite{KrohnFM04} to allow future file
requestors to perform order-independent verification on these data
blocks. Afterwards, according to $F$'s file identifier, the provider
$P$ maps the information of file $F$ (including the file identifier,
digest list and his own user identifier) onto the associated
maintainer $M$, i.e., $successor(F)$ (step 1 in
Figure~\ref{fig:stages}). Once the maintainer $M$ has received such
file information, he can additionally take hash of the entire digest
list as a \emph{version identifier} of file $F$. Moreover, in order
to support the social-based reputation model (described in
section~\ref{subsec:reputation}), each user who has voted some
versions of the file $F$ should inform the associated maintainer $M$
(step 1' in Figure~\ref{fig:stages}), so that the maintainer can
know which users have voted the versions of file $F$, i.e., the
associated \emph{voters}. In particular, since many providers may
individually publish a number of versions of file $F$, the
information\footnote{In this paper, we omit the metadata-related
information for clarity.} of $F$ stored at the maintainer $M$ is
shown in Table~\ref{tab:file}.

\begin{table}[tbp]
    \centering
    \caption{The information of a specific file}
    \label{tab:file}
    \begin{tabular}{|c|c|c|c|c|}
        \hline        & \textbf{Version} & \textbf{Digest} & \textbf{Provider ID} & \textbf{Voter ID} \\
                      & \textbf{ID}      & \textbf{List}   & \textbf{List}        & \textbf{List}     \\
        \hline    $1$ & $VI_1$           & $DL_1$          & $PIL_1$              & $OIL_1$           \\
        \hline    $2$ & $VI_2$           & $DL_2$          & $PIL_2$              & $OIL_2$           \\
        \hline \ldots & \ldots           & \ldots          & \ldots               & \ldots            \\
        \hline    $i$ & $VI_i$           & $DL_i$          & $PIL_i$              & $OIL_i$           \\
        \hline \ldots & \ldots           & \ldots          & \ldots               & \ldots            \\
        \hline    $n$ & $VI_n$           & $DL_n$          & $PIL_n$              & $OIL_n$           \\
        \hline
    \end{tabular}
\end{table}

However, a maintainer may be offline and a malicious or compromised
maintainer has the capacity of fabricating/refusing/corrupting the
file information maintained by himself. Therefore, we map the
information of a specific file $F$ onto $m$ maintainers
$\{M_i\}_{i=1}^{m}$ to provide the mechanisms of redundancy and
reliability, i.e., each file $F$ corresponds to a unique group of
maintainers.

\begin{equation} \label{eqn:multi-maintainers}
    M_i = successor(filename | i), \quad 1 \leq i \leq m
\end{equation}

\noindent Here, $M_i$ denotes the $i^{th}$ maintainer of the file
$F$, $filename$ denotes the name of the file, and $m$ denotes the
number of maintainers associated with the file. Due to the essential
feature of structured overlay, these $m$ associated maintainers are
distributed in the identifier circle uniformly at random; therefore,
we can design a filtering scheme to exclude the malicious
activities.

\subsection{Proactive Filtering}
\label{subsec:filtering}

Since the providers publish much information about their shared
files, we could collect such information from the associated
maintainers (step 2 in Figure~\ref{fig:stages}), and then provide a
\emph{proactive} defense against the content pollution (step 3 in
Figure~\ref{fig:stages}) as follows.

In Green, a requestor $R$, who wants to acquire a specific file $F$,
issues a query (i.e., $F$'s file identifier) towards the $m$
associated maintainers in the system according to the expression
\ref{eqn:multi-maintainers}. Each of these maintainers should
maintain the information of at least one version of the requested
file $F$. Then, these $m$ maintainers respond with the corresponding
information of the requested file $F$ (see Table \ref{tab:file}). On
harvesting these responses, the requestor $R$ mixes them and
generates a list $L$ consisting of $\langle VI, PI \rangle$ pairs
(do not remove duplicate pairs), where the $VI$ and $PI$ denote the
version and provider identifiers associated with the requested file
$F$, respectively.

Based on the generated list $L$, we design a \emph{filtering scheme}
to exclude the malicious activities. Without loss of generality, we
assume that the total percentage of malicious and compromised users
is $\beta$. Furthermore, as specified in
section~\ref{subsec:publication}, each file corresponds to a group
of $m$ maintainers. Due to the features of hash-based maintainer
assignment (see expression~\ref{eqn:multi-maintainers}), these
maintainers associated with the same file are distributed uniformly
at random. Therefore, the probability that $x$ out of $m$
maintainers in a group are malicious is

\begin{equation}
    \Pr(m, x) = {m \choose x} \times \beta^{x} \times (1 - \beta)^{m - x}
\end{equation}

\noindent Thus, the probability that less than half of these $m$
maintainers are malicious is

\begin{equation} \label{eqn:filtering}
    \Pr(m, 0, \lfloor \frac{m - 1}{2} \rfloor) = \sum_{x = 0}^{\lfloor \frac{m - 1}{2} \rfloor} \Pr(m, x)
\end{equation}

The expression~\ref{eqn:filtering} implies that though the malicious
and compromised maintainers may collectively create/delete/modify
their maintained $\langle VI, PI \rangle$ pairs, each $\langle VI,
PI \rangle$ pair existing in the generated list $L$ for no less than
$\lceil \frac{m + 1}{2} \rceil$ (i.e., $m - \lfloor \frac{m - 1}{2}
\rfloor$) times is authentic with the probability of $\Pr(m, 0,
\lfloor \frac{m - 1}{2} \rfloor)$. In practice, the system designer
would need to estimate the worst network environment that the system
should sustain in order to determine the corresponding
$\beta_{worst}$. Then, according to the $\beta_{worst}$ and the
expected communication overhead, the system designer could further
tune the parameter $m$ to guarantee that $\Pr(m, 0, \lfloor \frac{m
- 1}{2} \rfloor)$ is large enough even in the worst environment. For
instance, if $\beta_{worst} = 0.2$ and $m = 5$, the $\Pr(m, 0,
\lfloor \frac{m - 1}{2} \rfloor)$ will be $94.2\%$, which is
sufficient for common systems. In other words, by applying the
suitable system parameters, the requestor $R$ can safely employ a
\emph{proactive filtering} mechanism here by filtering the
aforementioned list $L$ through removing the $\langle VI, PI
\rangle$ pairs with less than $\lceil \frac{m + 1}{2} \rceil$ times.

In the same way, the requestor $R$ could filter out the malicious
$\langle VI, OI \rangle$ pairs, where the $VI$ and $OI$ denote the
version and voter identifiers associated with the requested file
$F$, respectively. Moreover, the malicious maintainers may choose to
corrupt a version identifier $VI_i$ or a digest list $DL_i$ of the
file, but this generally could not take effect since the requestor
is able to detect such corruption by simply hashing the digest list
and then comparing the result with the corresponding version
identifier. Certainly, these malicious and compromised maintainers
may choose to corrupt a version identifier $VI_i$ and the
corresponding digest list $DL_i$ simultaneously; if so, the
requestor could also utilize the above proactive filtering mechanism
to filter out the corrupted information. After all the previous
kinds of filtering, the requestor excludes these filtered
information, and groups the remainder to locally reconstruct the
information of the requested file $F$ (see Table~\ref{tab:file}).

\subsection{Social-Based Reputation Model}
\label{subsec:reputation}

Reputation model is a common technique to address the problem of
content pollution. In this section, we first design a basic
reputation model, and then we use the social information to enhance
its effectiveness and efficiency.

\subsubsection{Basic Reputation Model}
\label{subsubsec:basic}

Based on the proactive filtering mechanism, the requestor $R$ can
obtain the \emph{authentic} information of the requested file $F$ by
filtering out the malicious activities performed by malicious and
compromised maintainers. However, polluters could choose to simply
inject polluted versions of the file $F$ without corrupting the
maintained information. Due to the lack of authenticity
verification, the requestor $R$ generally cannot know which versions
have been polluted. Here, we resort to the general reputation-based
information to further help requestors identify the polluted
versions.

\textbf{Vote Gathering (step 4 in Figure~\ref{fig:stages}).} The
requestor $R$ first extracts each version $V_i$'s associated voter
identifier list $OIL_i$ from the reconstructed information of file
$F$ (see Table~\ref{tab:file}). Then, the requestor traverses
$OIL_i$, and contacts each voter $O_{ij}$ indicated by $OIL_i$ to
gain the voter's past vote history $OH_{ij}$. Once receiving these
vote histories, the requestor can easily obtain each voter
$O_{ij}$'s vote $o_{ij}$ on the version $V_i$. The vote may be
either $-1$, if the voter considers the version polluted, or $1$,
otherwise.

In the above simple vote gathering mechanism, voters are required to
remain online, otherwise their votes would be lost. To solve this
problem, besides the locally maintained vote history, we employ a
redundancy mechanism (similar to the
expression~\ref{eqn:multi-maintainers}) to map each user $U$'s vote
history onto a group of $m$ vote maintainers (steps 1'' and 1''' in
Figure~\ref{fig:stages}).

\begin{equation} \label{eqn:multi-vote-maintainers}
    OM_i = successor(UI | i), \quad 1 \leq i \leq m
\end{equation}

\noindent Here, $OM_i$ denotes the $i^{th}$ vote maintainer for the
user $U$, $UI$ denotes the identifier of the user, and $m$ denotes
the number of vote maintainers for the user. Now, no matter the
voter $O_{ij}$ is online or offline, the requestor $R$ has the
capacity of obtaining $O_{ij}$'s past vote history from the
associated vote maintainers, and then filtering out the malicious
activities performed by malicious and compromised vote maintainers
as described in section~\ref{subsec:filtering}. Note that, when the
number of a version's associated voters is too large, the requestor
should choose a set of randomly-selected voters to perform the vote
gathering in order to control the communication overhead.

\textbf{Reputation Computation (step 5 in Figure~\ref{fig:stages}).}
With the vote gathering mechanism, the requestor $R$ could obtain
each associated voter $O_{ij}$'s vote $o_{ij}$ on the version $V_i$.
Based on these votes, the requestor is able to compute the
reputation score of $V_i$. The simplest way is to execute the
unweighted averaging on these votes; however, it cannot distinguish
between different voters, e.g., both like-minded voters and
malicious voters are treated equally. Instead, in our design, we
compute a Credence-style~\cite{WalshS06} weight coefficient $\theta$
for each vote and execute the weighted averaging to compute the
reputation score. The computation of weight coefficient is described
as follows:

\begin{equation} \label{eqn:credence}
    \theta_{(R, O_{ij})} =
        \begin{cases}
            \frac{p - ab}{\sqrt{a(1-a)b(1-b)}} & \textrm{if $| \frac{p - ab}{\sqrt{a(1-a)b(1-b)}} | \geq 0.5$} \\
            0 & \textrm{if $|\frac{p - ab}{\sqrt{a(1-a)b(1-b)}}| < 0.5$}
        \end{cases}
\end{equation}

\noindent Here, given the set of versions on which both $R$ and
$O_{ij}$ have voted, $a$ and $b$ are the fraction where $R$ and
$O_{ij}$ voted positively, respectively, and similarly $p$ is the
fraction where both voted positively. Several abnormal cases that
arise in practice are discussed in~\cite{WalshS06}.

The weight coefficient $\theta_{(R, O_{ij})}$ expresses the
statistical correlation between the two users' vote histories, and
captures whether they tend to vote identically (i.e., $\theta_{(R,
O_{ij})} \geq 0.5$), inversely (i.e., $\theta_{(R, O_{ij})} \leq
-0.5$) or uncorrelatedly (i.e., $|\theta_{(R, O_{ij})}| < 0.5$).
Based on the weight coefficient, the requestor performs the weighted
averaging to compute the reputation score of each version $V_i$.

\begin{equation} \label{eqn:reputaion}
    Rep(V_i) = \frac{\sum_{j = 1}^{|OIL_i|} \left( o_{ij} \times \theta_{(R, O_{ij})} \right) }{\sum_{j = 1}^{|OIL_i|} | \theta_{(R,
    O_{ij})} | } \in [-1, 1]
\end{equation}

\noindent Here, $|OIL_i|$ denotes the size of the associated voter
identifier list of $V_i$; moreover, if $\sum_{j = 1}^{|OIL_i|} |
\theta_{(R, O_{ij})} | = 0$, then $Rep(V_i) = 0$. This weighted
averaging scheme gives more weight to votes from like-minded voters,
and it can be used to assist requestors in better identifying
polluted versions.

\subsubsection{Social-Based Enhancement}
\label{subsubsec:enhancement}

Currently, many P2P systems have already merged the \emph{social}
information, e.g., the \emph{friend} links, for some purposes. These
friend links are very different from the overlay links, and they
could be created in various ways. For instance, the friends may be
these real-world acquaintances, or many more. A user and his friends
generally share a similar interest and give similar votes on a
specific version; moreover, the friends are much more trustworthy
than other common users. By exploiting the information of friends,
we can provide two kinds of enhancement for the above basic
reputation model: \emph{vote history extension} and \emph{efficiency
improvement}.

\textbf{Vote History Extension.} In a P2P content sharing system, a
requestor may have only a few local vote history. This would make
the requestor cannot accurately compute the correlation between each
associated voter and himself, thus influencing the performance of
our reputation model. To extend the local vote history, the
requestor could additionally consider his friends' vote histories
before performing the reputation computation.

We assume that the requestor $R$ with local vote history $OH_R$ has
$f$ friends in the system, denoted by $\{F_i\}_{i=1}^{f}$; moreover,
each friend $F_i$ has the vote history $OH_{F_i}$. In our vote
history extension scheme, the requestor first collects his friends'
vote histories from the associated vote maintainers, by filtering
out the malicious activities performed by malicious and compromised
vote maintainers. Then, the requestor adopts an attenuation
coefficient $\gamma$ to weigh his friends' vote histories. Finally,
the requestor performs a simple \emph{nonzero} averaging on these
friends' vote histories as well as his local vote history to compute
the \emph{extended} vote history $OH'_R$.

\begin{equation} \label{eqn:extension}
    \begin{aligned}
        OH'_R & = avg \left( OH_R, OH_{F_1} \times \gamma, \cdots, OH_{F_f} \times \gamma \right) \\
              & = avg \left( OH_R, \{ OH_{F_i} \times \gamma \}_{i=1}^{f} \right)
    \end{aligned}
\end{equation}

\noindent Here, each vote history is treated as a vector, and the
$avg$ is defined to be a function of computing, in turn, the
averages of \emph{nonzero} values at each position in these vectors.

To illustrate the computation process intuitively, we give an
example. Suppose a requestor $R$ with local vote history $OH_R = \{
1, -1, 0, 0, -1, 0 \}$ has two friends $F_1$ and $F_2$ in the
system. So, he is able to collect the two friends' vote histories
$OH_{F_1}$ and $OH_{F_2}$ from the associated vote maintainers.

\begin{center}
    \begin{tabular}{l @{ } l @{ } l @{} r}
        $OH_{F_1}$ & $=$ & $\{ 0, 0, 1, 0, -1,$ & $-1 \}$\\
        $OH_{F_2}$ & $=$ & $\{ 1, 0, 1, 0, -1,$ & $ 0 \}$
    \end{tabular}
\end{center}

\noindent In practice, the vote history can be stored and
transmitted in a more compact way. If the attenuation coefficient
$\gamma$ is set to $0.9$, then the final extended vote history
$OH'_R$ is

\begin{center}
    \begin{tabular}{l @{ } l @{ } l @{} l}
        $OH'_R$ & $=$ & $\{ \frac{1 + 1 \times 0.9}{1 + 0.9},$ & $\frac{-1}{1}, \frac{1 \times 0.9 + 1 \times 0.9}{0.9 + 0.9}, 0,$\\
                &     &                                        & $\frac{(-1) + (-1) \times 0.9 + (-1) \times 0.9}{1 + 0.9 + 0.9}, \frac{(-1) \times 0.9}{0.9} \}$\\
                & $=$ & $\{ 1, -1, 1,$                         & $0, -1, -1 \}$
    \end{tabular}
\end{center}

Due to the fact that the friends are generally trustworthy, the
requestor could apply the vote history extension to enrich his vote
history reliably; therefore, he is able to perform the reputation
computation more accurately. Note that, if a requestor $R$ has
\emph{many} malicious friends, the extended vote history may be
polluted; here, some existing mechanisms, e.g.,
SybilGuard~\cite{YuKGF06} and SybilLimit~\cite{YuGKX08}, could be
used to improve the trustworthiness of friends. Moreover, if the
requestor does not have any friends in the system, our social-based
reputation model falls back to the basic form. In some sense, the
requestor should ``pay the price'' for having no friends or having
many malicious friends.

Interestingly, our proposed vote history extension scheme can not
only improve the accuracy of reputation computation, but also solve
the ``cold start'' problem, i.e., a newly incoming user (i.e.,
newcomer) without any vote history cannot benefit from the
reputation model. To address the ``cold start'' problem, once
joining the system, the newcomer first builds up the friend links.
Then, he collects his friends' vote histories to compute the
extended vote history as described above. Specifically, this
extended vote history could be treated as the newcomer's initial
local vote history, so that he could perform the reputation model
normally.

\textbf{Efficiency Improvement.} From the previous description of
vote history extension, each time a requestor executes the
reputation computation, he should collect the vote histories from
his friends' associated vote maintainers, which is somewhat
expensive. In Green, each user \emph{periodically} collects his
friends' vote histories from these associated vote maintainers, and
meanwhile, each user maintains a \emph{friend-vote} database to
store the latest collected vote histories (step 1'''' in
Figure~\ref{fig:stages}). With this kind of database, the requestor
is able to retrieve his friends' vote histories locally.

Secondly, due to the fact that a user and his friends generally have
the similar interest, they may give the similar votes on a specific
version. If many friends of a requestor have already voted a
version, the requestor does not need to compute the reputation score
of the version again; as an alternative, he could utilize these
friends' votes to efficiently estimate the reputation score.

Without loss of generality, we assume that the requestor wants to
obtain the reputation score of a specific version $V_i$, and
moreover he has $f$ friends, among which there are $f'$ friends
$\{F_j\}_{j=1}^{f'}$ who have already given the votes
$\{o_{ij}\}_{j=1}^{f'}$ on the version $V_i$. Note that, such
information can be retrieved from the locally maintained friend-vote
database. Specifically, if only a few friends have voted the version
(i.e., $f'$ is small), the associated votes may be biased, so now we
have to return back to use the normal social-based reputation model
described before; otherwise, if a sufficient number ($f' \geq 4$ in
our design) of friends have voted the version, we utilize the
associated votes to efficiently estimate the reputation score
$Rep'(V_i)$ of the version $V_i$.

\begin{equation} \label{eqn:quick-evaluate}
    Rep'(V_i) = \frac{ \sum_{j=1}^{f'} o_{ij} }{ f' } \times \gamma
\end{equation}

\noindent Here, the parameter $\gamma$ denotes the attenuation
coefficient. If $| Rep'(V_i) | \geq 0.5$, we consider that these
friends voted the version $V_i$ identically, so the $Rep'(V_i)$ can
be treated as the final reputation score $Rep(V_i)$ of the version
$V_i$; otherwise, there are significant differences among these $f'$
associated friends, so we again return back to use the normal
social-based reputation model described before.

With low probability, a malicious user may masquerade as genuine
user to become the requestor's friend, and a friend may also be
compromised. To address this ``malicious friend'' problem, before
performing the efficient reputation estimation, the requestor should
first compute the correlation coefficient $\theta$ between each
associated friend and himself, as specified in
expression~\ref{eqn:credence}. If $\theta < 0.5$, the friend may be
malicious or uncorrelated~\cite{WalshS06}, so we choose not to take
this friend's vote into account. Due to the periodically updated
friend-vote database, these correlation coefficients could be
computed periodically instead of repeatedly.

In summary, we use the social (friend-related) information \emph{a)}
to provide a vote history extension mechanism to enhance the
effectiveness of basic reputation model even for a system newcomer,
and \emph{b)} to further enhance the efficiency of both the vote
history collection and reputation computation.

\subsection{Version Selection}
\label{subsec:select}

After the proactive filtering (section~\ref{subsec:filtering}) and
social-based reputation computation
(section~\ref{subsec:reputation}), the requestor has obtained two
kinds of information about each version $V_i$ of his requested file:
the authentic provider identifier list $PIL_i$ and the reputation
score $Rep(V_i)$. Based on such information, the requestor is able
to select a version for downloading (step 6 in
Figure~\ref{fig:stages}).

Since the size of the associated provider identifier list generally
reflects the availability of a specific version, in many cases, the
requestor $R$ selects the version with the largest provider
identifier list. However, the analytical model presented
in~\cite{DumitriuKKSZ05} characterized the impact of various
selection strategies, and implied that the ``largest'' strategy is
vulnerable to the pollution attack as well as the Denial-of-Service
attack. As an alternative, in Green, the requestor $R$'s choice is
biased towards versions with more providers. Furthermore, the
reputation score of a version indicates the authenticity of the
version, so the probability of selecting a version should also be
proportional to the associated reputation score. Considering the
above factors, we get the following expression:

\begin{equation} \label{eqn:selection}
    \Pr(V_i) = \frac{|PIL_i| \times \widetilde{Rep(V_i)}}{\sum_{j = 1}^{n} \left( |PIL_j| \times \widetilde{Rep(V_j)} \right)}, \quad 1 \leq i \leq n
\end{equation}

\noindent Here, $\Pr(V_i)$ denotes the probability of selecting the
version $V_i$ for downloading, $|PIL_i|$ denotes the size of the
provider identifier list of $V_i$, and $n$ denotes the number of
versions associated with the requested file. Specifically, as the
original $Rep(V_i) \in [-1, 1]$, we make a linear mapping and let
$\widetilde{Rep(V_i)} = Rep(V_i) + 1$ to ensure each $\Pr(V_i) \geq
0$. Note that, when the denominator of the expression is $0$, the
requestor has to perform an ad hoc version selection; however, this
might happen only when all versions' original reputation scores are
$-1$, which indicates that there is no authentic version of the
requested file existing in the system.

\subsection{Downloading with Realtime Verification}
\label{subsec:downloading}

The requestor $R$ starts downloading the data blocks of the selected
version from these associated providers in parallel. Once the
requestor receives a data block, he first locally computes a digest
$D$ by hashing the received data block, and then, matches $D$
against the digest existing in the corresponding digest list of the
requested $F$'s file information (reconstructed based on the
proactive filtering as described in section~\ref{subsec:filtering}).
If they are matched, the received data block is accepted, otherwise
the block is dropped. These operations make the requestor have the
capacity of verifying the integrity of the selected version while
downloading (step 7 in Figure~\ref{fig:stages}).

In common as described above, aiming at verifying the data
integrity, the requestor $R$ verifies all the received data blocks
while downloading. To reduce the verification overhead, we propose a
\emph{probabilistic verification scheme} which verifies randomly
selected blocks instead of all blocks.

In this scheme, the requestor $R$ assumes that a polluter should
tamper with at least $r$ blocks of all $b$ blocks, and moreover,
less than $r$ polluted blocks could be successfully recovered by the
error-correcting code (ECC) technique. Based on this assumption, the
requestor $R$ would need to determine the \emph{expected false
positive rate (EFPR)} of the probabilistic verification according to
the tradeoff between integrity assurance and verification overhead.
That is,

\begin{equation} \label{eqn:fpr}
    \begin{aligned}
        FPR & =
            \begin{cases}
                \frac{{b - r \choose v}}{{b \choose v}} = \frac{(b-r)! \times (b-v)!}{(b-r-v)! \times {b!}} & \textrm{if $r+v \leq b$} \\
                0 & \textrm{if $r+v>b$}
            \end{cases}\\
        & \leq EFPR
    \end{aligned}
\end{equation}

\noindent Here, $FPR$ and $EFPR$ denote the actual and expected
false positive rate, respectively; moreover, $b$ denotes the total
number of data blocks, $v$ denotes the number of verified data
blocks, and $r$ denotes the lower bound of the number of polluted
data blocks. Now, the requestor $R$ can randomly verify the minimal
number of data blocks, denoted by $v_{min}$, that satisfies the
above expression \ref{eqn:fpr}, in order to reduce the verification
overhead.

Finally, after the downloading and verification, the requestor
builds the complete version from these data blocks. If the built
version is authentic, the requestor gives/publishes a positive vote
on the version, and then shares the downloaded version; otherwise,
if the version is unfortunately found to be polluted, the requestor
should delete it, give/publish a negative vote, and then repeat from
the step of version selection (section~\ref{subsec:select}) until he
receives the authentic version or the authentic version is found to
be non-existent in the system.

\subsection{Deployment in Unstructured Overlay}
\label{subsec:unstructured}

In the previous sections, we focus on the design of Green in
structured overlay systems; to illustrate the broad applicability,
we further describe how to deploy Green in unstructured overlay
systems.

Generally, a unstructured overlay system is composed of users
joining the overlay with some loose rules. To deploy our design, we
introduce a two-tier structure into the unstructured overlay: a
subset of users (i.e., ultra-users) construct a \emph{structured}
mesh, while the other users (i.e., leaf-users) connect to the
ultra-user tier for participating into the overlay system.
Specifically, we adopt the election algorithm in~\cite{LuaZ07} to
select the ultra-users, and then utilize our proposed design to
construct the structured mesh among ultra-users, so that the
unstructured overlay systems could also benefit from our defense
mechanisms against content pollution. In practice, many modern
unstructured P2P overlay systems have already implemented some
similar two-tier structures, so that the deployment of Green in
unstructured overlay systems will not make many modifications to
their original network structures. Consequently, Green is broadly
applicable, and it is able to work for both structured and
unstructured overlays.

\section{System Analysis}
\label{sec:analyze}

A well-designed pollution-free P2P content sharing system should
seek to optimize its defense capacity and maintenance overhead under
the two kinds of common content pollution attacks discussed in
section~\ref{sec:attack}.

\subsection{Defense Capacity}
\label{subsec:capacity_analysis}

\textbf{Decoy Insertion.} Under a decoy insertion attack, the
polluters inject a massive number of corrupted versions (with the
same metadata but different identifiers) for a target file, in order
to reduce the availability of authentic versions of the file. In
Green, we propose a social-based reputation model to defend against
such decoy insertion.

In a P2P content sharing system, the genuine users give positive
votes on the authentic versions, and negative votes on the polluted
versions; on the contrary, in order to spread the polluted versions,
the polluters give the \emph{opposite} votes to value the polluted
versions instead of authentic versions. However, this kind of
malicious voting operations will notably impact the correlation
between polluters and genuine users (see
expression~\ref{eqn:credence}), and make genuine users give negative
weights ($\theta < 0$) to the votes from polluters, i.e., a negative
vote from polluters actually values a version (see
expression~\ref{eqn:reputaion}). Even if the polluters apply a
\emph{tricky} voting strategy by giving correct votes sometimes,
this will also not influence the system performance significantly
because the genuine users would mark the votes from these tricky
polluters as uncorrelated and then directly drop them. Moreover,
each requestor considers his friends' votes to extend the vote
history reliably (see expression~\ref{eqn:extension}), so that the
requestor could perform the reputation computation more accurately.
To enhance the efficiency, if many friends of a requestor have voted
a specific version, the requestor further utilizes his friends'
votes to quickly estimate the reputation score of the version (see
expression~\ref{eqn:quick-evaluate}).

Consequently, though polluters inject many corrupted versions of a
target file into the system and perform the malicious voting, the
genuine users are still able to select the authentic versions for
downloading effectively and efficiently.

\textbf{Identifier Corruption.} To launch an identifier corruption
attack, the polluters should first analyze the properties of the
hash function used by a P2P content sharing system. Then, they
accurately corrupt a fraction of data (e.g., the data that are not
used) without changing the identifier of the corrupted version.
Unfortunately, the reputation model cannot identify such attacks due
to the unchanged version identifier. In our design, we propose a
block-based verification mechanism to defend against the identifier
corruption.

In Green, a maintainer manages many digest lists, each of which is
associated with a maintained version. A requestor could obtain the
digest lists, and then utilize them to verify the integrity of data
blocks while downloading (see section~\ref{subsec:downloading}). Due
to the fact that each data block is used to generate the digest list
and each block digest is verified, it is very difficult for
polluters to perform the identifier corruption attack. Furthermore,
we design a probabilistic verification mechanism to reduce the
verification overhead. Here, each requestor \emph{randomly} verifies
a number of data blocks of the selected version (see
expression~\ref{eqn:fpr}), so that the polluters cannot always
launch successful identifier corruption attacks by corrupting some
\emph{fixed} data blocks. Indeed, the probabilistic verification may
impact our defense capacity against identifier corruption. There is
a tradeoff between integrity assurance and verification overhead. If
a requestor wants to completely defeat the identifier corruption, he
should verify all the downloaded data blocks.

\subsection{Maintenance Overhead}

To support the running of Green, a participating user plays two
roles: file information maintainer and vote maintainer; and
meanwhile, each user concretely maintains three information sets: a
set of file information, a set of common users' vote histories, and
a set of his friends' vote histories. Specifically, in Green, the
maintained information is distributed averagely, and it will not
incur much maintenance overhead on the participating users. The
reason is twofold.

Firstly, since we use the hash function to perform the assignment of
file information maintainers, each user maintains the information of
roughly the same number of files, thus achieving a good load
balancing. As shown in Table~\ref{tab:file}, the information of a
file is composed of a number of $\langle VI_i, DL_i, PIL_i, OIL_i
\rangle$ quaternions. In each quaternion, the version identifier
$VI_i$ is a hash value; the digest list $DL_i$ is a list of data
block digests, here the system designer could control the size of
$DL_i$ through bounding the number of divided data blocks of a file;
moreover, both $PIL_i$ and $OIL_i$ merely store the associated
provider identifiers and voter identifiers, respectively, which will
not aggravate a user's maintenance burden significantly. In
addition, even if a user maintains the information of a popular file
(with a large number of versions, providers and voters), several
techniques~\cite{GodfreyLSKS04,RaoLSKS03} could be employed here for
load balancing.

Secondly, due to the hash-based vote maintainer assignment and the
redundancy mechanism, each user maintains roughly $m$ common users'
vote histories on average; furthermore, each user maintains a
friend-vote database consisting of his friends' vote histories.
Generally, maintaining these vote histories will also not aggravate
the burden of a user significantly.

\section{Evaluation} \label{sec:evaluation}

In this section, we first describe the experimental setup, and then
we present the key performance metric. Finally, we comparably
evaluate the performance of Green.

\subsection{Experimental Setup}

To evaluate the performance, we developed a prototype system
implementing all our proposed security mechanisms with approximately
$5040$ lines of Java code. Specifically, all the experiments are run
on an OpenPower720 with four-way dual-core POWER5 CPUs and 16GB RAM
running SLES 9.1.

\textbf{Network Model:} We use two realistic massive-scale network
traces~\cite{MisloveMGDB07}, YouTube (1,157,827 users and 4,945,382
friend links, Jan 2007) and Flickr (1,846,198 users and 22,613,981
friend links, Jan 2007), to generate the social/friend networks.
Specifically, since these traces do not indicate which users are
genuine or malicious, we have to perform an appropriate processing
on the two generated social networks.

According to the small-world property of online social
networks~\cite{AhnHKMJ07,MisloveMGDB07}, we generate $50$ genuine
user groups for each generated social network as follows. We first
randomly select $50$ bootstrapping users as the seeds of these
genuine user groups, and then use the breadth-first search algorithm
to extend these genuine user groups to guarantee that each group
contains a different number of \emph{unique} users (Zipf
distribution with its parameter $\alpha = 1$). Note that, if we
cannot generate a required number of genuine users from a
bootstrapping user, we randomly select another bootstrapping user to
replace this one. Except these users contained by genuine user
groups, the other users are polluters.

We choose Chord~\cite{StoicaMKKB01} as the underlying P2P overlay of
Green, and all the users participate in the Chord overlay to
construct a structured overlay network, so that we could use it to
assign/lookup the file information maintainers, the vote
maintainers, etc.

\textbf{Content Model:} The previous study in~\cite{LiangKXR05}
reported the existence of a large number of polluted versions for a
single file. In our experiments, there are $10,000$ unique files,
each of which has $100$ different versions. Since we do not model
the corruption activities performed by malicious maintainers, the
information of a file is stored at only one information maintainer,
and similarly the vote history of a user is also stored at one vote
maintainer. Moreover, each version is associated with a number of
providers which vary over time.

At the start of our experiments, each genuine user shares $20$
authentic versions, and each polluter shares $400$ and $50$ polluted
versions when performing decoy insertion and identifier corruption,
respectively; this models a highly malicious environment. Here, the
versions shared by a user are determined by first selecting a
certain file and then its version. Specifically, both selections
follow Zipf distribution with $\alpha = 0.8$~\cite{LiangKXR05}.

\textbf{Execution Model:} Different queries are initiated at
uniformly distributed users, and the attenuation coefficient
$\gamma$ is set to $0.9$. Specifically, an experiment is composed of
$50$ experimental cycles, and each experimental cycle is divided
into $5,000,000$ query cycles. In each query cycle, the selection of
a specific version to download is done by first selecting a file
according to Zipf distribution with $\alpha = 0.8$, and then
choosing a version based on our proposed schemes. After each
experimental cycle, the number of authentic downloads is computed.

For a genuine user, he gives a vote on the downloaded version with a
probability of $P_o$. Here, we add noise to model users' mistakes
through making genuine users vote correctly with only $90\%$
probability. Further, if a downloaded version is polluted, the
genuine user deletes this version with a probability of $P_d$. On
the other hand, for a polluter, he always gives a malicious (i.e.,
opposite) vote on the downloaded version, and then shares it. In
addition, each user collects his friends' vote histories from the
associated vote maintainers and updates the local friend-vote
database once per $100,000$ query cycles.

\subsection{Performance Metric}

In our experiments, we characterize the system performance using the
\emph{fraction of authentic downloads}. It is defined as the
fraction of downloads that the genuine users acquire authentic
versions during one experimental cycle.

\subsection{Evaluation Results}

\begin{figure*}[tbp]
    \centering
    \subfloat[Comparison using YouTube trace]
    {
        \label{fig:decoy_comparison_youtube}
        \begin{minipage}[t]{0.245\textwidth}
            \centering
            \includegraphics[width=\textwidth]{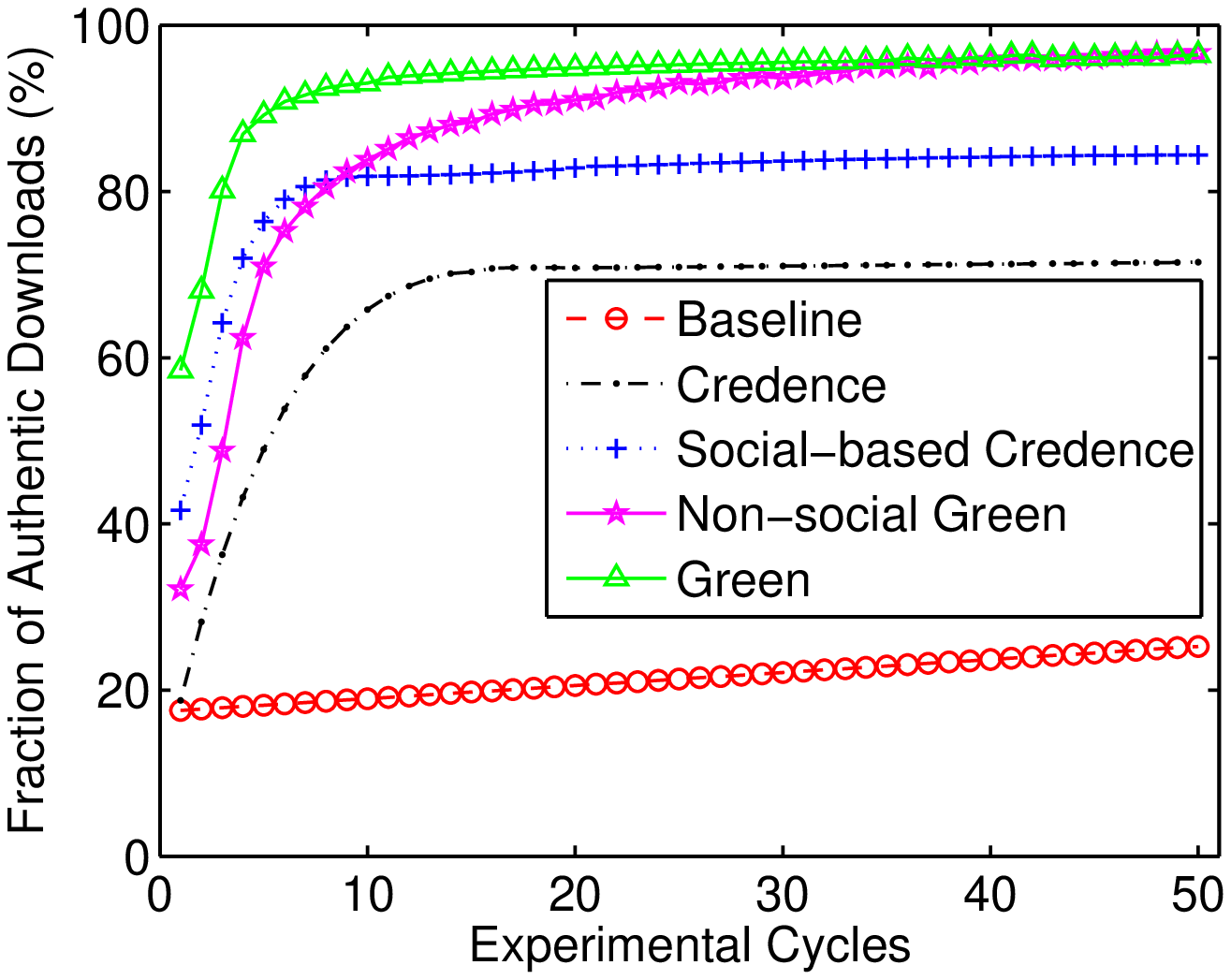}
        \end{minipage}
    }
    \subfloat[Comparison using Flickr trace]
    {
        \label{fig:decoy_comparison_flickr}
        \begin{minipage}[t]{0.245\textwidth}
            \centering
            \includegraphics[width=\textwidth]{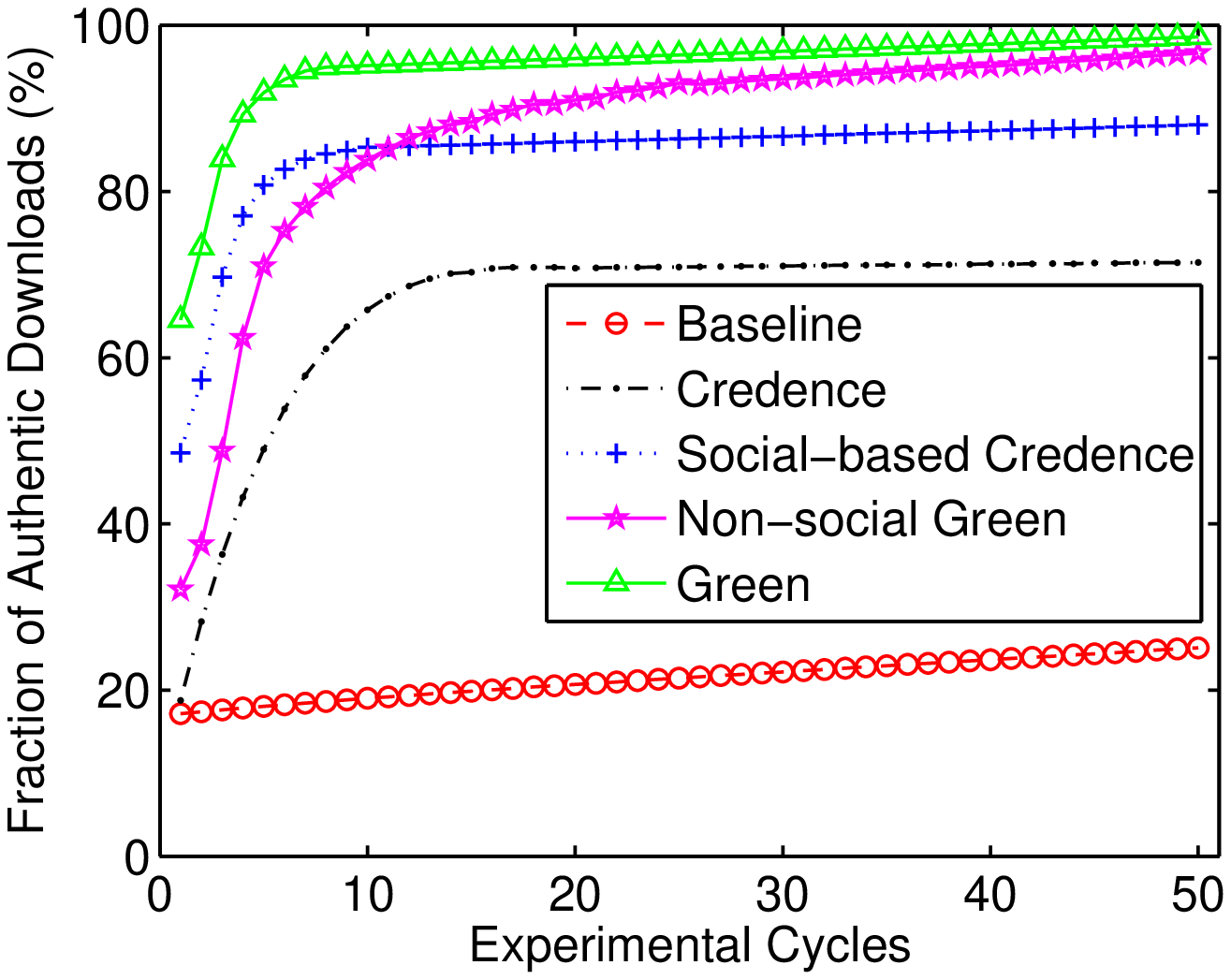}
        \end{minipage}
    }
    \subfloat[Impact of polluters]
    {
        \label{fig:decoy_polluters_ours}
        \begin{minipage}[t]{0.245\textwidth}
            \centering
            \includegraphics[width=\textwidth]{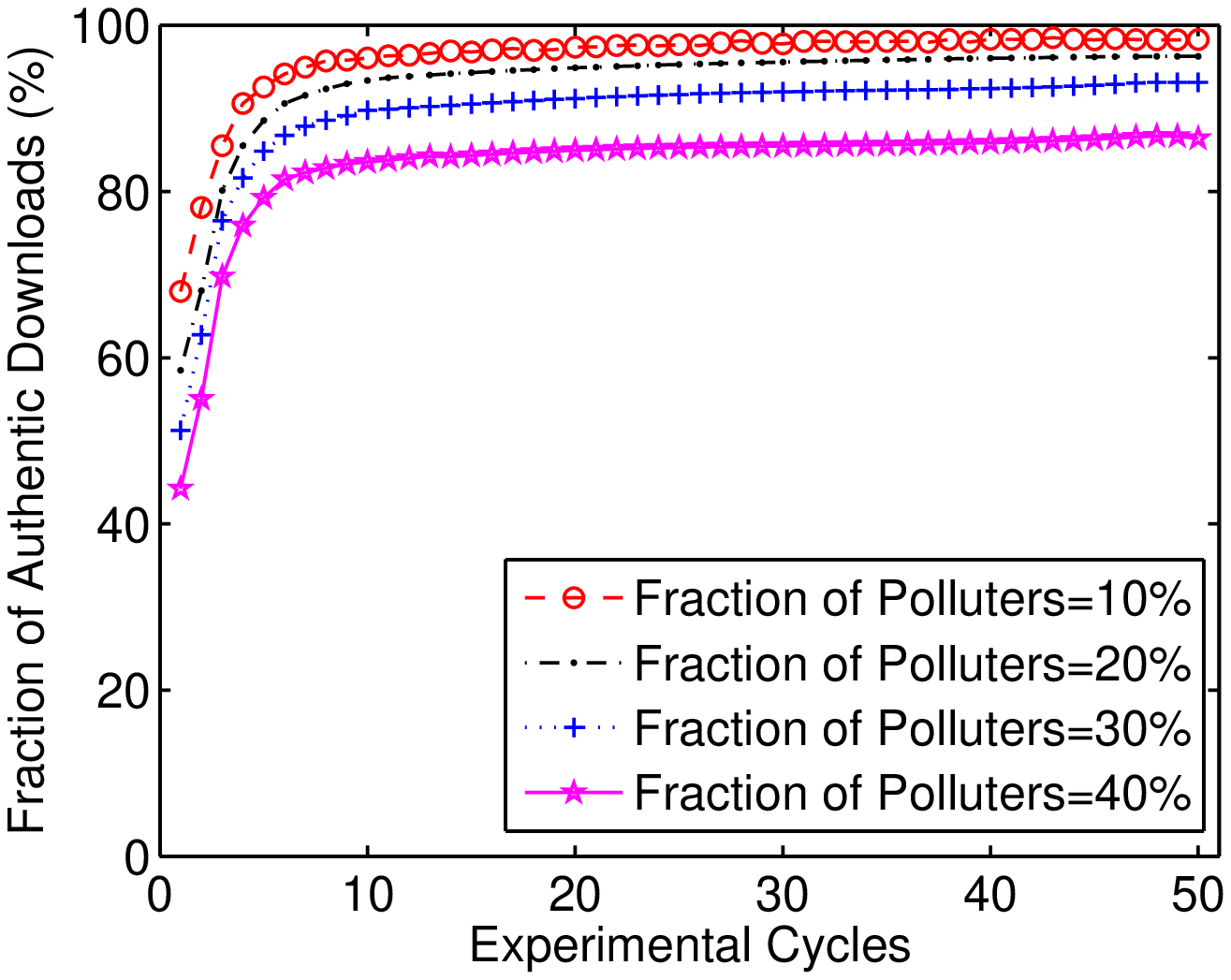}
        \end{minipage}
    }
    \subfloat[Impact of voting]
    {
        \label{fig:decoy_voting_ours}
        \begin{minipage}[t]{0.245\textwidth}
            \centering
            \includegraphics[width=\textwidth]{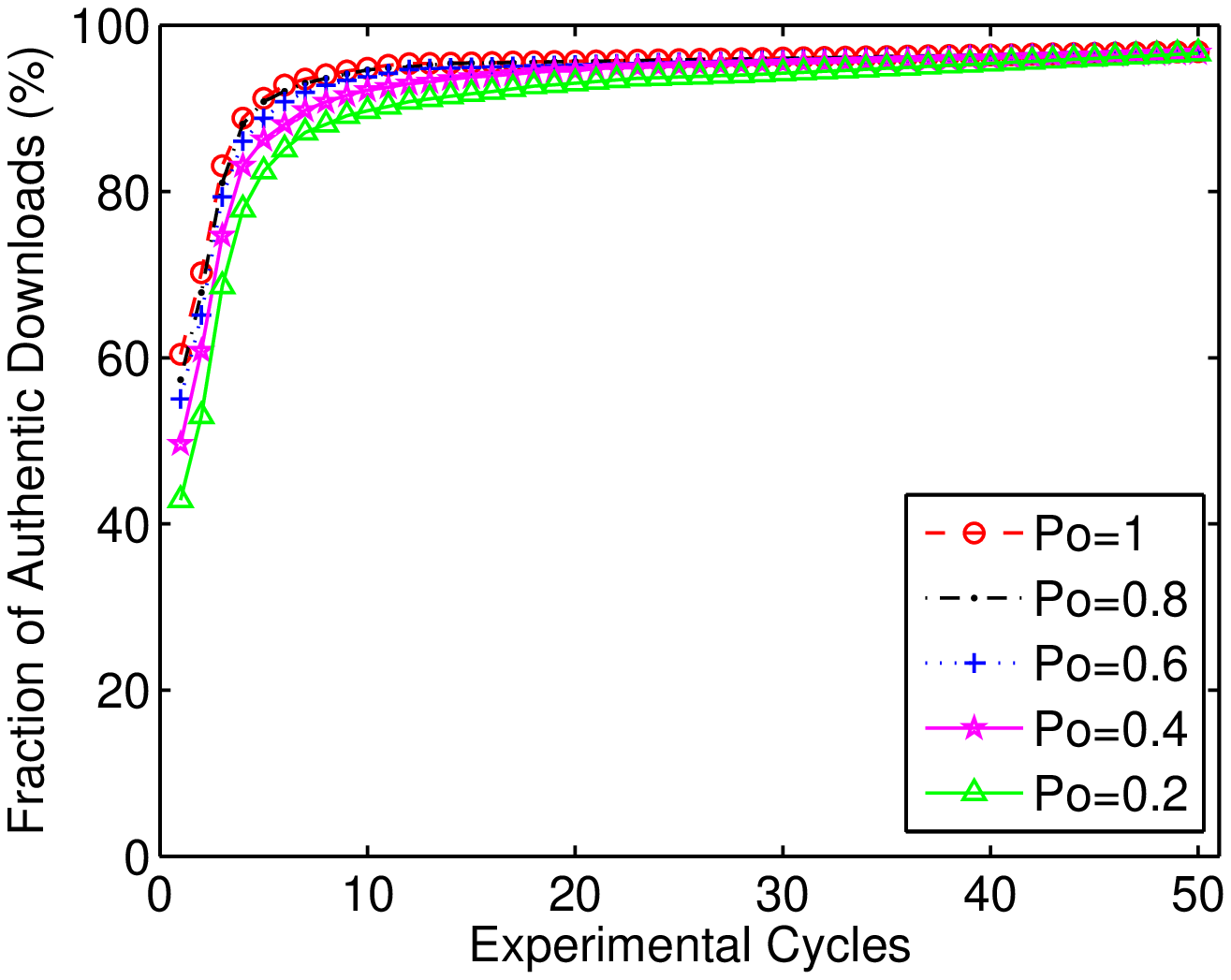}
        \end{minipage}
    }
    \caption{Evaluation results under the decoy insertion attack}
    \label{fig:decoy_insertion}
\end{figure*}

In this section, we evaluate the performance of Green under decoy
insertion and identifier corruption, by considering four parameters:
$\beta$, $P_o$, $P_d$ and $FPR$ (see Table~\ref{tab:exp_notation}).
Their default values are $0.2$, $0.9$, $0.9$ and $0.133$,
respectively. Specifically, in each experiment, we may vary only one
parameter and leave the others default.

\begin{table}[tbp]
    \centering
    \caption{Experimental Notation}
    \label{tab:exp_notation}
    \begin{tabular}{|r|l|}
        \hline \textbf{Symbol} & \textbf{Meaning} \\
        \hline $\beta$         & The total fraction of polluters \\
               $P_o$           & The probability that a genuine user \\
                               & gives a vote on the downloaded version \\
               $P_d$           & The probability that a genuine user \\
                               & deletes the downloaded polluted version \\
               $FPR$           & The false positive rate of probabilistic \\
                               & verification \\
        \hline
    \end{tabular}
\end{table}

\subsubsection{Under Decoy Insertion Attack}
\label{subsubsec:decoy_evaluation}

\textbf{Comparison.} Using two realistic traces, we evaluate the
performance of Green as compared to four other systems with the
default parameters.

\begin{itemize}
    \item Baseline: The probability of selecting a version $V_i$ to download is proportional to $|PIL_i|$ (i.e., the number of $V_i$'s associated providers).
    \item Credence: An ingenious and representative pollution defense system deployed on live networks~\cite{WalshS06}.
    \item Social-based Credence: The Credence system with our social-based enhancement.
    \item Non-social Green: The Green system without our social-based enhancement.
\end{itemize}

As shown in Figures~\ref{fig:decoy_comparison_youtube} and
\ref{fig:decoy_comparison_flickr}, Green outperforms all other
systems with only $3\%$ of all the downloaded versions being
polluted; moreover, the ``Baseline'' cannot work well. Specifically,
two systems with social-based enhancement, i.e., Green and
``Social-based Credence'', converge faster than ``Non-social Green''
and ``Credence''. This is because that, though each user has only a
small number of vote history at the startup, the user in
social-based systems could additionally consider his friends' vote
histories to help judging the version authenticity more accurately.

Furthermore, we could find that the two systems with social-based
enhancement, i.e., Green and ``social-based Credence'', in the
network with Flickr trace converge \emph{a little} faster than in
the network with YouTube trace. The reason is that each user in
Flickr has $12.25$ friend links on average which is larger that
$4.27$ in YouTube. This phenomenon also implies that, in order to
achieve a good performance, a user in Green actually does not need
too many friends.

Since Green always outperforms the four other systems in the
following experiments and it performs similarly in the two networks
with YouTube and Flickr traces, respectively, we hereafter choose to
merely present the experimental results of Green in the network with
YouTube trace for saving the space.

\textbf{Impact of Polluters.} In this experiment, we investigate the
influence incurred by different fractions of polluters ($\beta$). As
shown in Figure~\ref{fig:decoy_polluters_ours}, the performance of
Green decreases slightly with the growth of the fraction of
polluters, and Green can work well even in a highly malicious
environment with $40\%$ of users being polluters. Here, the slight
performance decrease is mostly due to the fact that, with the
increase of polluters, both the fraction of genuine users' malicious
friends and the number of polluted versions increase in the system.
Also, the result implies that if a user could ensure most of his
friends being genuine, he is able to select an authentic version to
download with very high probability; this actually provides a strong
incentive for genuine users to maintain the trustworthiness of their
friends.

\begin{figure*}[tbp]
    \centering
    \subfloat[Comparison using YouTube trace]
    {
        \label{fig:id_comparison_youtube}
        \begin{minipage}[t]{0.245\textwidth}
            \centering
            \includegraphics[width=\textwidth]{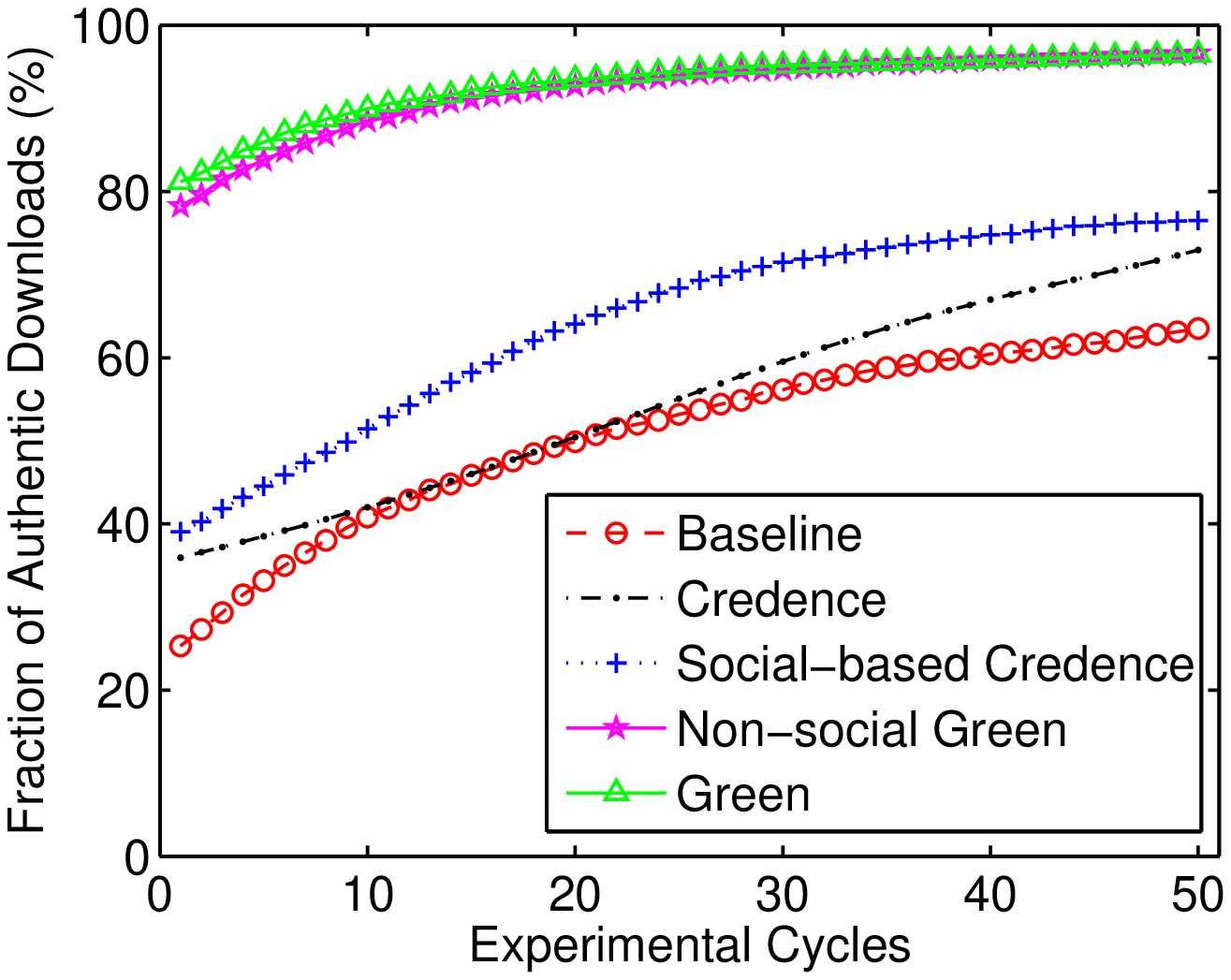}
        \end{minipage}
    }
    \subfloat[Impact of polluters]
    {
        \label{fig:id_polluters_ours}
        \begin{minipage}[t]{0.245\textwidth}
            \centering
            \includegraphics[width=\textwidth]{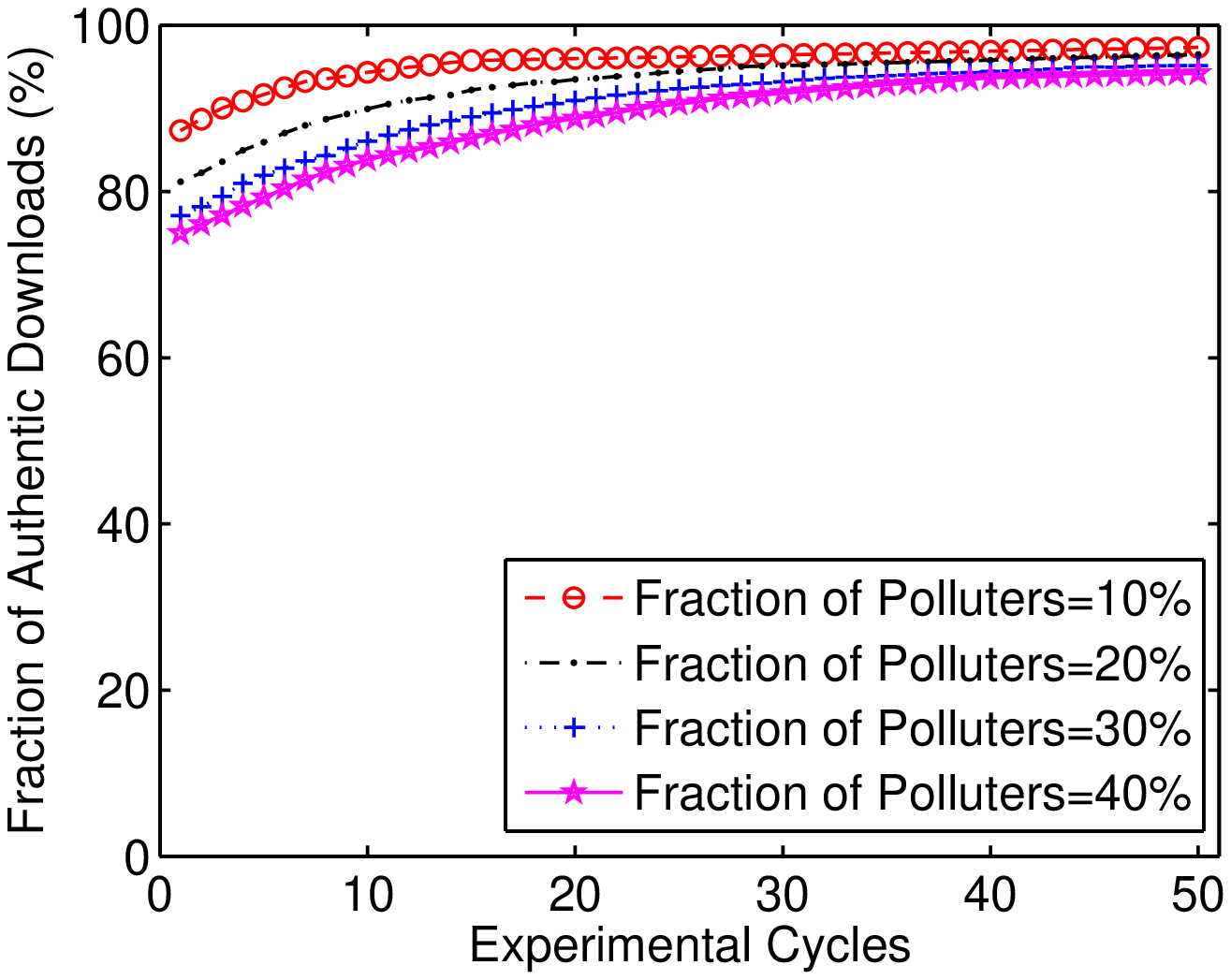}
        \end{minipage}
    }
    \subfloat[Impact of voting]
    {
        \label{fig:id_voting_ours}
        \begin{minipage}[t]{0.245\textwidth}
            \centering
            \includegraphics[width=\textwidth]{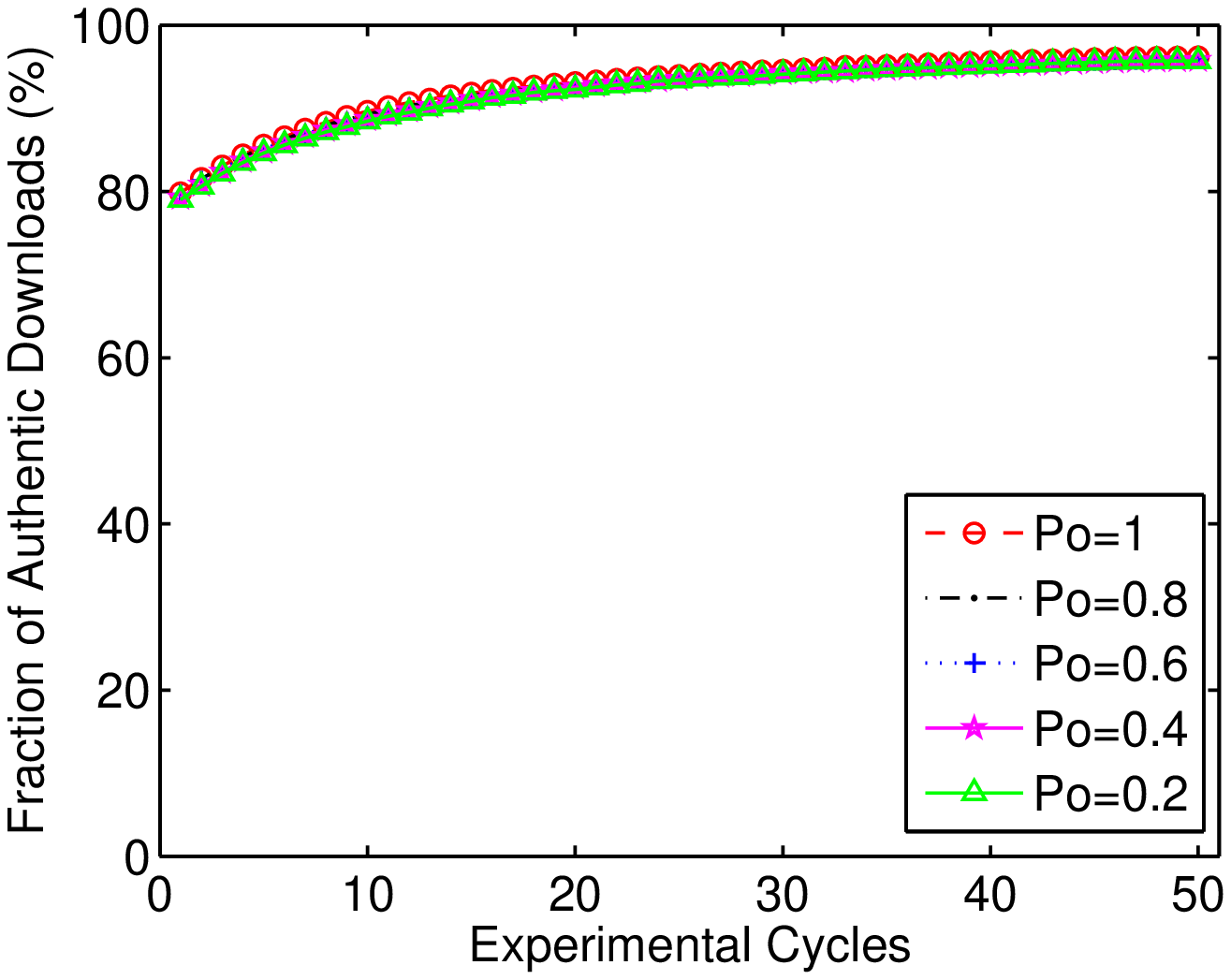}
        \end{minipage}
    }
    \subfloat[Impact of FPR]
    {
        \label{fig:id_fpr_ours}
        \begin{minipage}[b]{0.245\textwidth}
            \centering
            \includegraphics[width=\textwidth]{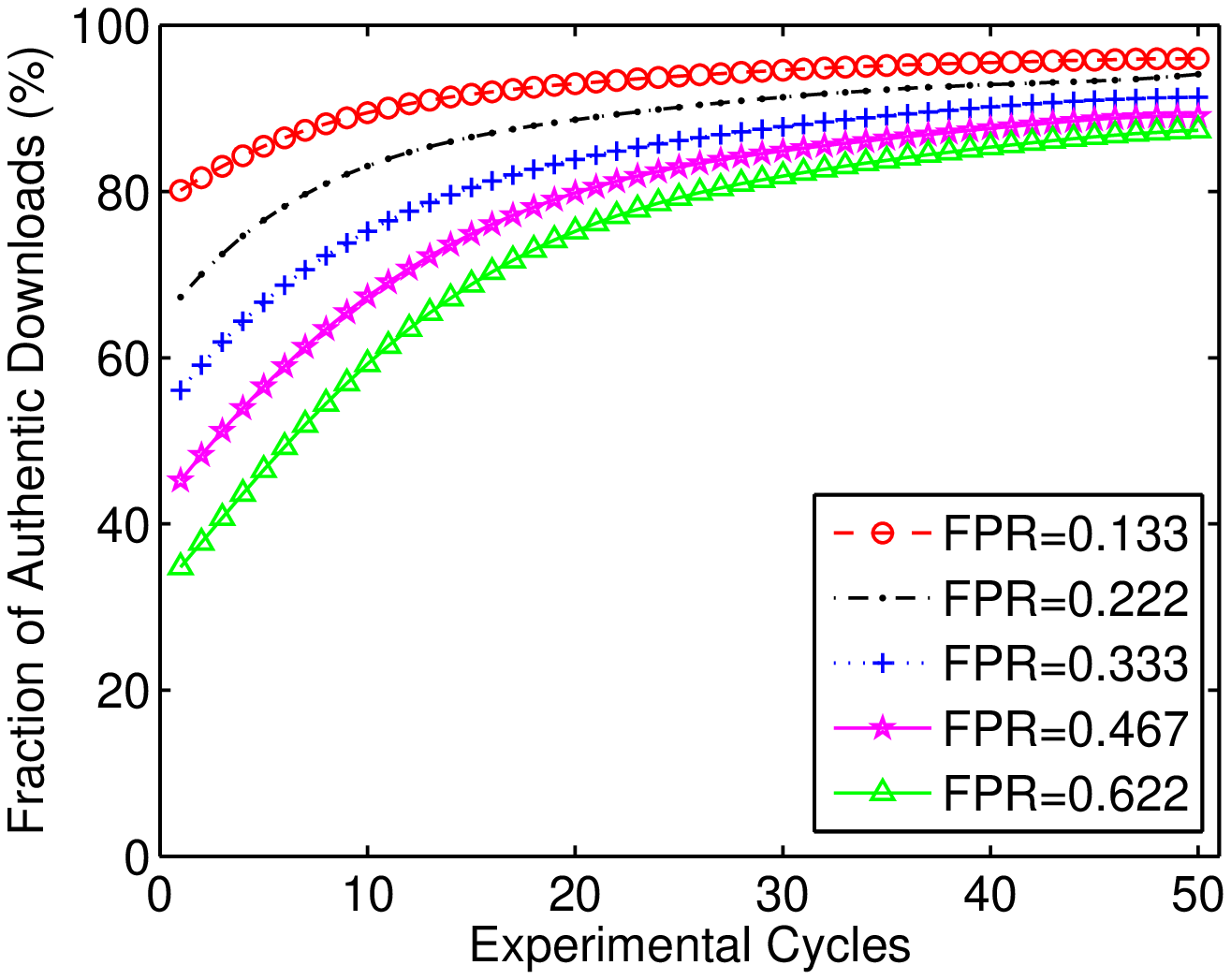}
        \end{minipage}
    }
    \caption{Evaluation results under the identifier corruption attack}
    \label{fig:decoy_insertion}
\end{figure*}

\textbf{Impact of Voting.} Figure~\ref{fig:decoy_voting_ours} shows
that Green can work well with various different voting probabilities
($P_o$); moveover, even if genuine users are very inactive to vote
the downloaded versions ($P_o = 0.2$), Green can still work well.
This benefits from our vote history extension scheme which could
extend a user's vote history by considering his friends' vote
histories. Specifically, in current P2P content sharing systems,
only a few users may vote the downloaded versions~\cite{LiangKXR05},
so the above experimental result further indicates that Green can
indeed be deployed in practice. Note that, if $P_o = 0$, Green would
fall back to a Baseline-like system and any existing reputation
systems cannot work, under decoy insertion.

\textbf{Impact of Deleting.} We further evaluate the system
performance with different deleting probabilities ($P_d$). In Green,
the deleting strategy impacts the number of a polluted version's
associated providers, thus influencing the probability of selecting
this version to download in the future (see
expression~\ref{eqn:selection}). Hence, we would expect that the
system performance increases with the growth of deleting
probability.

Interestingly, different deleting probabilities have, however, no
significant impact on the performance of Green. Our analysis takes
three factors into consideration. Firstly, a user may activate his
own deleting module only after he has already downloaded a polluted
version. Secondly, Green is able to identify most of the polluted
versions, and therefore, a genuine user downloads polluted versions
with very low probability. Thirdly, even if a genuine user has
downloaded a polluted version and does not delete it, he may give a
negative vote on this version; thus, other genuine users could
identify this polluted version based on the negative vote. As a
result, no matter which deleting strategy is adopted by these
genuine users, Green could always work well under decoy insertion
attacks. Specifically, Green also keeps this property under
identifier corruption attacks. Due to the space limit, we omit the
two corresponding figures.

\subsubsection{Under Identifier Corruption Attack}

\textbf{Comparison.} In this experiment, we comparably evaluate the
performance of Green in the network with YouTube trace.
Figure~\ref{fig:id_comparison_youtube} illustrates that Green
outperforms all other four systems. Moreover, the three systems,
i.e., ``Baseline'', ``Credence'' and ``Social-based Credence'',
which do not perform verification while downloading cannot defend
against identifier corruption effectively; this validates the
effectiveness of our proposed realtime verification mechanism. In
addition, the experimental result also indicates that the effect of
social-based enhancement is limited under the identifier corruption
attack.

Due to the space limit and the similar performance in both networks
with the YouTube and Flickr traces, respectively, we omit to present
the system performance using Flickr trace, and hereafter choose to
merely present the experimental results of Green in the network with
YouTube trace.

\textbf{Impact of Polluters.} We vary the fraction of polluters
($\beta$), and investigate the influence incurred by polluters.
Figure~\ref{fig:id_polluters_ours} shows that Green can defeat
identifier corruption even in a highly malicious environment with
$40\%$ of users being polluters. In real-world P2P systems, some
polluters may perform Sybil~\cite{Douceur02} attacks to create a
large number of virtual malicious users, and the above experimental
result implies that Green could also work well under the Sybil
attack. Of course, we are able to incorporate Green with some other
Sybil defenses, e.g., SybilGuard~\cite{YuKGF06} and
SybilLimit~\cite{YuGKX08}, to achieve a better system performance.

\textbf{Impact of Voting.} As shown in
Figure~\ref{fig:id_voting_ours}, we evaluate the system performance
with the voting probability ($P_o$) varying in steps of $0.2$. The
experimental result illustrates that the voting probability has
little effect on the system performance, i.e., our system always
works well no matter which voting strategy the genuine users adopt.
The reason of this phenomenon is that Green's defense capacity
against identifier corruption is mostly provided by the realtime
verification mechanism which does not utilize the vote history. This
experimental result further validates that our proposed mechanism is
applicable in real-world P2P systems where only a few users may give
votes on the downloaded versions.

\textbf{Impact of False Positive Rate.} In Green, each user is able
to flexibly adjust the false positive rate ($FPR$) of probabilistic
verification. The experimental result in
Figure~\ref{fig:id_fpr_ours} shows that the false positive rate of
probabilistic verification impacts our defense capacity against the
identifier corruption attack. There is a tradeoff between integrity
assurance and verification overhead. The more data blocks a user
verifies, the better defense capacity he can gain; whereas the
pattern is reversed. This result validates our analysis in
section~\ref{subsec:capacity_analysis}.

\section{Related Work} \label{sec:related}

\subsection{Pollution Defenses}

So far, many reputation models have been proposed to address the
problem of content pollution in P2P content sharing systems. In
general, these reputation models can be grouped into three main
categories: \emph{peer-based} models, \emph{object-based} models and
\emph{hybrid} models.

In peer-based reputation models, e.g., EigenTrust~\cite{KamvarSG03},
PeerTrust~\cite{XiongL04} and Scrubber~\cite{CostaSAA07}, to reflect
the level of honesty, each participating user is assigned a
reputation score by considering his past behaviors in pairwise
transactions. According to the reputation score, genuine users could
collectively identify content polluters, and then isolate these
polluters from the system. However, the studies
in~\cite{DumitriuKKSZ05,WalshS06} evaluated the potential impact of
peer-based reputation models, and implied that these models are
insufficient to defend against the pollution attack.

Among the object-based reputation models, Credence~\cite{WalshS06}
is the typical representative of them. In Credence, genuine users
determine the object authenticity through secure tabulation and
management of endorsements from other users. This model utilizes the
statistical correlation to measure the reliability of users' past
votes, and designs a decentralized flow-based trust computation to
discover trustworthy users. However, a newcomer without vote history
hardly has any capacity of distinguishing between authentic objects
and polluted objects.

Aiming at combining the benefits of both peer-based and object-based
models, several hybrid reputation models, e.g.,
XRep~\cite{DamianiVPSV02}, X$^2$Rep~\cite{CurtisSS04} and extended
Scrubber~\cite{CostaA07}, have been further presented. Nevertheless,
due to the fact that most of the participating users in P2P content
sharing systems are rational in seeking to maximize their individual
utilities, the reputation models are penalized by the lack of
reliable user cooperation.

Currently, without resorting to reputation models, several other
pollution defenses have been proposed. Micropayment techniques,
e.g., MojoNation~\cite{MojoNation} and PPay~\cite{YangG03}, can be
utilized to counter the pollution attack by imposing a cost on
content polluters --- to inject polluted content into the system
they should first commit a certain amount of resources. Furthermore,
Habib \emph{et al.} in~\cite{HabibXABC05} developed an integrated
security framework to verify content integrity in P2P media
streaming. This framework could provide high assurance of data
integrity with low computation and communication overheads; however,
it requires centralized supplying peers. In~\cite{MichalakisSG07},
Michalakis \emph{et al.} presented a ``Repeat and Compare'' system
to ensure content integrity by utilizing the P2P substrate to repeat
content generation on other peers and compare the results to detect
content pollution. This system focuses on the replication of
computations instead of data replication and comparison, and its
application field is generally the P2P content distribution
networks. Recently, Chen \emph{et al.} in~\cite{ChenLCGTC08}
proposed a new pollution defense based on the notion that the
content providers are the only sources to accurately distinguish
polluted contents and verify the integrity of the requested
contents; however, it cannot defeat decoy insertion effectively.

\subsection{Social Networking}

Social networking sites such as YouTube, Facebook, MySpace, Orkut
and Flickr are experiencing explosive growth, both in terms of
involved communities and overall participating users. The social
networking technique has significantly impacted how Internet users
make use of the today's Internet. So far, several proposals have
been made to improve existing distributed systems by exploiting the
inherent properties of social networks.

PGP~\cite{Zimmermann95} is one of the early social networking
applications, in which the participants create a ``web of trust'' to
authenticate public keys based on their acquaintances' opinions in a
fully self-organized manner. This ``web of trust'' model utilized
the friend-of-friend trust structure, and then Garriss \emph{et al.}
in~\cite{GarrissKFKMY06} adopted the similar structure to develop
the ``Reliable Email (R{\footnotesize E}:)'', an automated
whitelisting-based email acceptance system, to mitigate email spam.
R{\footnotesize E}: exploits social relationships among email
correspondents, and moreover, it does not incur false positives
among socially connected users. In~\cite{MisloveGD06}, Mislove
\emph{et al.} tried to combine the information contained in both
hyperlinks and social links, so they merged the social networking
technique into the Web search engine to optimize the ranking results
by considering various interests of different users.

To defend against Sybil attacks~\cite{Douceur02}, two promising
decentralized social-based protocols, SybilGuard~\cite{YuKGF06} and
SybilLimit~\cite{YuGKX08}, have been proposed. They utilize the
social networks among user identities based on the fact that Sybil
users could create many identities but few trust relationships;
finally, the two protocols have the capacity of allowing only very
limited Sybil users to be accepted even in a large-scale network.
Furthermore in~\cite{MislovePDG08}, based on the similar fact that
it is difficult for a user to create arbitrarily many trust
relationships, the Ostra system explored the use of existing social
links to impose a cost on the information senders, thus preventing
the adversary from sending excessive unwanted communication.
Recently, Ramachandran and Feamster in~\cite{RamachandranF08}
proposed a framework called Authenticatr to establish authenticated
out-of-brand communication channels between applications by
utilizing social links existing in various social networking sites.

\section{Conclusion} \label{sec:conclusion}

In this paper, we design a pollution-free P2P content sharing
system, \emph{Green}. In Green, a content provider (i.e., creator or
sharer) publishes the information of his shared contents into a
security overlay; in order to obtain the authentic contents, a
content requestor can utilize the inherent content-based information
to perform the proactive filtering and realtime verification, and
utilize the traditional reputation information and the social
information to perform the social-based reputation computation.
Green is broadly applicable for both structured and unstructured
overlay applications. The analysis and prototype-based evaluation
(in massive-scale networks with realistic traces) indicate that
Green is able to defend against the content pollution effectively
and efficiently.


{\footnotesize \bibliographystyle{acm}
\bibliography{draft}}

\end{document}